\documentclass[11pt,twoside]{article}
\usepackage{asp2004}
\usepackage{psfig}
\usepackage{epsf}
\usepackage{graphics}
\usepackage{lscape}
\def\edcomment#1{\iffalse\marginpar{\raggedright\sl#1\/}\else\relax\fi}
\reversemarginpar

\begin{document}
\title{Correlated X-ray and Optical Variability in X-ray Binaries}
\author{Robert I. Hynes}
\affil{The Louisiana State University, Department of Physics and
  Astronomy, Baton Rouge, Louisiana, USA; \\The University of Texas at
  Austin, McDonald Observatory and Department of Astronomy, 1
  University Station C1400, Austin, Texas 78712-0259, USA;\\
  rih@phys.lsu.edu; Hubble Fellow}

\begin{abstract}
It has long been known that X-ray and optical/UV variability in active
X-ray binaries is sometimes correlated.  The expectation has been that
this arises from rapid reprocessing of X-rays into longer wavelength
photons, and hence that the lags between the bandpasses can be used to
reconstruct an echo-map of the binary, revealing the geometry and
spatial scale of reprocessing sites.  I will review how this can be
done, what can be learned, and what progress has been made
with real data.  In addition, an interesting twist to the story has
emerged in the form of correlated variability that does not seem
associated with reprocessing.  This extends to very short timescales,
includes quasi-periodic behavior, and may be associated with optical
synchrotron emission from a jet.  Finally, contrary to expectations we
have recently discovered that similar correlations are seen even in an
ostensibly `quiescent' object allowing similar analyses of
reprocessing effects.
\end{abstract}
\thispagestyle{plain}

\section{Introduction}

X-ray binaries, and more specifically the Low Mass X-ray Binaries
(LMXBs) that are the focus of this review, are variable objects.  This
is a universal characteristic seen from the radio through optical to
X-rays and $\gamma$-rays.  The X-ray variability is usually dominated
by instabilities in the accretion flow.  X-rays irradiate the outer
accretion disk and companion star, resulting in reprocessed optical
and UV radiation which is expected to be imprinted with the same
variability as the X-ray signal is.  An important difference, however,
is that the optical and X-rays originate from a volume of significant
spatial extent, resulting in light travel time delays between the
X-rays and the reprocessed emission.  It is then possible to infer
information about the geometry and scale of the reprocessing region
from the lags measured between X-ray and optical/UV variability; this
technique is known as reverberation or echo-mapping, as the
reprocessed light behaves as an echo.

In this work we will review the concepts underlying the echo-mapping
technique and the successes achieved in applying it to LMXBs to date.
Surprisingly the technique is applicable not only in the most X-ray
luminous systems but in at least one {\em quiescent} black hole as
well.  In attempting to apply the technique it has also emerged that
not all optical/UV correlations arise in reprocessing at all.  We will
briefly consider an alternative kind of correlations that may
originate in optical/UV synchrotron emission from a jet.


\section{The Echo-Mapping Technique}

Echo or reverberation mapping are not uniquely applied to X-ray
binaries.  Much of the development and application of the technique
has been for active galactic nuclei (AGN); see for example
\citet{Peterson:2005a} for a recent review of the AGN problem and
\citet{OBrien:2002a} for the application to X-ray binaries.
The key idea is that optical (or UV) variability, in either lines or
continuum, is induced by reprocessing of X-ray variability, but lagged
by light travel times within the system.  Each reprocessing point can
be thought of as responding to X-rays with a $\delta$ function
response at a delay time determined by the path difference between
direct and reprocessed emission.  The total optical response is then
the sum of lagged responses from all the reprocessing elements.  For a
$delta$ function variation in the X-rays the optical response is then
termed the transfer function, and measures
how strong the response is as a function of the delay, effectively
encoding information about the reprocessing geometry.  For real
lightcurves, the optical lightcurve can be modeled as a convolution of
the X-ray lightcurve with the transfer function.

Several assumptions are inherent in this description.  It is assumed
that the optical responds linearly to the X-rays, or at least that a
non-linear response can be linearized for small perturbations.  It is
also implicit that the X-rays originate from a point source, or at
least a region much smaller in spatial extent than the reprocessing
region.  Finally to determine geometric information it is necessary
that the lags be geometric in origin; significant reprocessing times
would compromise this, and will be discussed in the next section.

\subsection{Geometrical modeling of the response}

\begin{figure}[t]
\begin{center}
\psfig{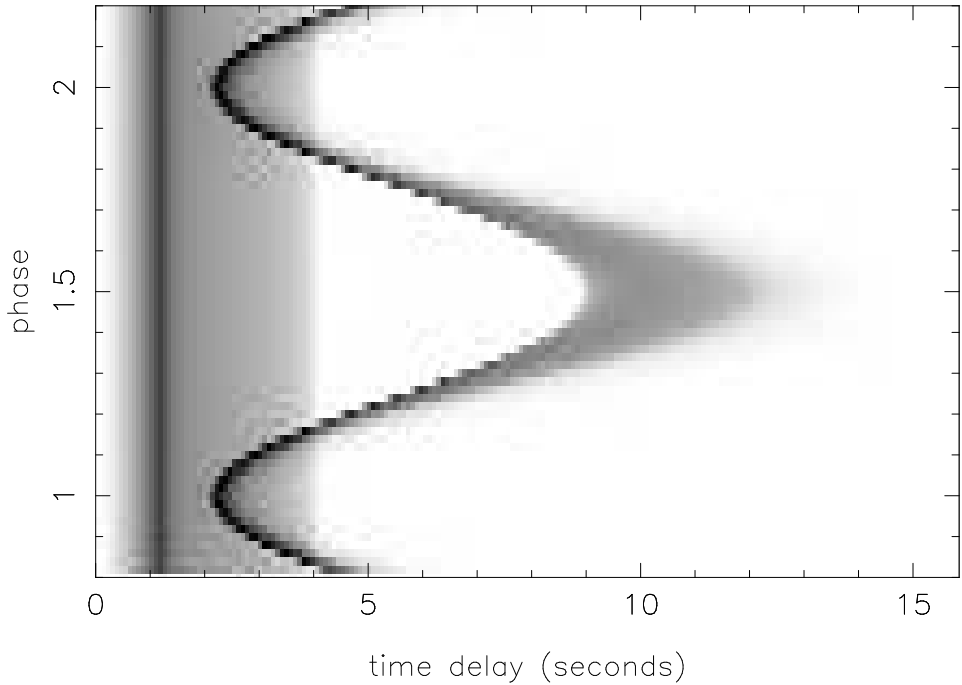}
\end{center}
\caption{Strength of response (indicated by grayscale) as a function
  of orbital phase and lag, as calculated by the binary modeling code
  of \citet{OBrien:2002a}.}
\label{EchoTomFig}
\end{figure}

In the case of X-ray binaries we have a clearer expectation of the
reprocessing geometry than in AGN.  We anticipate reprocessing from
the accretion disk around the compact object, possibly enhanced at a
bulge where material feeds into the disk from the companion star.  We
also might expect some reprocessing from the heated inner face of the
companion star.  \citet{OBrien:2002a} modeled the reprocessing
geometry to predict transfer functions for a variety of binary
parameters.  An example as a function of orbital phase is shown in
Fig.~\ref{EchoTomFig}.  Simplistically one expects two components.
The disk will extend from zero lag to $r_{\rm disk}(1 + \sin i)$ where
$r_{\rm disk}$ is the disk radius in light seconds and $i$ the binary
inclination.  Within this range the shape of the response is strongly
sensitive to the inclination and somewhat less so to the degree of
disk flaring.  The response from the companion star approximately
oscillates within the range $a(1 \pm \sin i)$ over the course of the
binary orbit, where $a$ is the binary separation in light seconds.
The strength and width of the companion response is a strong function
of the mass ratio and disk thickness (which determines how much of the
companion is shielded).  One of the great appeals of applying
echo-mapping to X-ray binaries is that with phase-resolved
observations of the companion echo over the orbit, one could measure
both $a$ and $i$ independently of other techniques and assumptions.

\subsection{Empirical descriptions of the response}

\citet{OBrien:2002a} did apply their model to real data, and we will
discuss their results in due course.  Many echo-mapping studies use
more pragmatic, and less model-dependent approaches.  Ideally one
would take high quality X-ray and optical lightcurves and deconvolve
them to directly determine the shape of the transfer function.  This
is the basis of the maximum entropy echo-mapping technique
\citep{Horne:1994a} in which a maximum entropy regularization method
is used to suppress the problem of fitting the noise.  Unfortunately
for many X-ray binary datasets, this technique has proved of limited
value.  This is because of the much shorter timescales involved than
in AGN, resulting in lower signal-to-noise data.  Typically one finds
that the transfer function is not well constrained in detail, and the
response one recovers is very sensitive to the assumptions made.

An alternative and simpler approach has thus been developed in which a
very simple functional form is adopted for the transfer function,
either a rectangular response \citep{Pedersen:1982a} or a Gaussian
\citep{Hynes:1998a}.  Both are introduced as approximations to the
response rather than for a physically motivated region, and amount to
the assumption that the data only constrain the mean lag and the
amount of smearing.  The Gaussian formulation essentially yields the
first two moments of the delay distribution.

An even simpler, but widely used approach is to measure the
cross-correlation of the two datasets.  This yields a mean lag but no
information about the smearing, and considerable care should be taken
in interpreting the results of a cross-correlation analysis
\citep{Koen:2003a}. 


\section{Reprocessing physics}

\subsection{Energy dependent reprocessing}

The response of an atmosphere, whether that of the disk or companion
star, is highly sensitive to the energy of incident photons, and hence
to the irradiating spectrum.  The primary energy absorption mechanism
will be photoabsorption, by K or L shell electrons of carbon,
nitrogen, or oxygen below the iron edge ($\sim7$\,keV), and by iron
above that.  Photoabsorption cross-sections are a steep function of
energy, so low energy photons will be absorbed in the upper layers of
the atmosphere, while higher energies will penetrate progressively
deeper.  Above $\sim10$\,keV photoabsorption cross-sections become
small enough that Compton scattering begins to take over as the
dominant source of X-ray opacity.  Some of the higher
energy photons will diffuse deeper into the atmosphere due to
scattering, but others will be scattered back out depositing only a
small fraction of their energy.  At higher energies this Compton
reflection becomes increasingly likely resulting in a high X-ray
albedo and low reprocessing efficiency.

These effects divide the response into three regimes.  At less than a
few keV, photons are photo-absorbed at low optical depths and can be
expected to produce an X-ray heated corona above the disk.
The reprocessed spectrum will not be efficiently thermalized and may
include a significant emission line component.  At intermediate
energies, up to $\sim10$\,keV, photons penetrate deeper into the
photosphere before being absorbed, resulting in absorption at moderate
optical depths.  In this case reprocessed optical and UV photons will
undergo multiple scatterings before diffusing out of the atmosphere
and a thermalized spectrum, closer to a black body, is expected.
Finally above 10\,keV, photons will increasingly tend to be Compton
reflected rather than photo-absorbed, and less efficient reprocessing
occurs.  

It is worth remarking that type-I X-ray bursts, to be discussed in
more detail subsequently, provide an ideal X-ray spectrum to observe
thermal reprocessing.  The typical 1--3\,keV black body spectrum
outputs most of its energy in the intermediate regime described above,
resulting in an efficient thermalized response.  Softer photons can be
expected to produce some lines as well, and harder ones to result in a
Compton reflected burst (see \citealt{Ballantyne:2004a} and references
therein.)

\subsection{Reprocessing times}

We have thus far discussed only lags due to global light travel times
within the system.  Local delays in reprocessing the X-rays might also
be expected, from two sources.  The first are diffusion times,
essentially local turbulent light travel times as reprocessed photons
undergo a random walk to escape from the atmosphere.  The second are
finite times associated with each interaction of a photon with an atom
or ion.

Diffusion times are expected to be the dominant of the two.  These
were briefly considered in the specific case of X-ray bursts by
\citet{Pedersen:1982a} who estimated that the {\em typical} diffusion
time will always be less than 0.6\,s, small compared to light travel
times within the binary.  \citet{Cominsky:1987a} examined the problem
more rigorously, calculating time-dependent responses of a hot stellar
atmosphere to an X-ray burst, including the effect of the burst on the
atmospheric temperature structure and opacities.  Their results were
in agreement with those of \citet{Pedersen:1982a}, and they found that
50\,\%\ of the reprocessed light is expected within just 0.2\,s, but
that there was also a very extended tail to the response up to 10\,s.
Their calculations also considered harder irradiation and cooler
atmospheres, finding that both would increase the diffusion timescale
significantly (see also \citet{McGowan:2003a}).  Thus while
reprocessing times are of marginal significance in considering burst
reprocessing, they may be of more importance for other applications
where further investigation is needed.


\section{Type I X-ray Bursts}

\subsection{Historical observations}

Type-I X-ray bursts are thermonuclear explosions on the surface of a
neutron star in an LMXB \citep{Strohmayer:2004a}.  They represent an
enormous increase in the X-ray flux, a factor of twenty or more,
rising on a timescale of a few seconds.  From these characteristics,
together with the optimal spectrum as already discussed, it should be
clear that X-ray bursts are an ideal echo-mapping probe.  Hence it is
using these events that some of the first echo-mapping experiments
were performed, and the most convincing results have been obtained.

Reprocessed optical bursts were discovered in the late 1970's in the
LMXBs 4U~1735--444 and Ser X-1 (\citealt{Grindlay:1978a};
\citealt{McClintock:1979a}; \citealt{Hackwell:1979a}).  The optical
flux was found to rise by nearly a factor of two and lag a few seconds
behind the X-rays.  It was immediately appreciated that the optical
flux was several orders of magnitude to high to be due to direct
emission from the neutron star surface, and hence that the brightening
must be due to reprocessing of X-rays by the much larger projected
area of the accretion disk and/or companion star.  The 2.8\,s lag in
4U~1735--444 \citep{McClintock:1979a} supported this interpretation,
being consistent with the expected light travel time delays
in this short-period binary.

Subsequent efforts focused on a number of coordinated campaigns to
observe another LMXB 4U~1636--536 (\citealt{Pedersen:1982a};
\citealt{Lawrence:1983a}; \citealt{Matsuoka:1984a};
\citealt{Turner:1985a}; \citealt{Truemper:1985a}).  These campaigns
yielded a total of twelve simultaneous bursts from this system.  Lags
were found to be in the range 0--4\,s with evidence that the variation
in the lags was a real effect and not due to measurement error
\citep{Matsuoka:1984a}.  In one burst \citep{Truemper:1985a} the
optical rise-time was clearly longer than that of the X-rays,
suggesting significant light-travel time smearing of a few seconds,
but in most cases the data quality were insufficient, allowing only an
upper limit on the smearing timescale.  

The large amplitudes of X-ray bursts provide additional information
not available when variability is a small perturbation.  Observations
never record the bolometric luminosity, but always a bandpass-limited
one.  Consequently the observed reprocessed lightcurve depends on the
spectral evolution as the reprocessor cools.  Shorter wavelengths are
sensitive to hotter material, they are expected to decay more rapidly,
and hence multicolor observations provide some temperature
sensitivity.  This is not a subtle effect, and the early observations
indicate the reprocessor temperature typically doubles during a burst.
The best considered case was analyzed by \citet{Lawrence:1983a}, who
found the reprocessor temperature rising from 25,000\,K to 50\,000\,K
at the peak of outburst.  All of the early observations were
constrained not only by data quality, however, but by the limitations
of optical data ($UBV$ photometry at best) in constraining such high
temperatures precisely.

In the time since these studies were performed, substantial steps
forward have been made both in X-ray sensitivity, most notably by {\it
RXTE}, and in the quality of optical data.  The replacement of
photomultipliers with CCDs was initially a hindrance to this work, as
high time-resolutions were not available, but the advent of fast CCD
systems has allowed enormous improvements in optical data quality
exploiting higher quantum efficiencies and two-dimensional detectors.
This is exemplified by simultaneous observations of a burst in
GS~1826--24 \citep{Kong:2000a}.  The higher quality of these
observations allowed determination of both a lag of $\sim3$\,s and a
dispersion of $\sim3$\,s, in spite of the source being fainter than
4U~1636--536 at both X-ray and optical wavelengths.  These
observations make it clear that the accretion disk must play a major
part in reprocessing of X-ray bursts, as might be expected.  The large
dispersions seen by \citet{Truemper:1985a} and \citet{Kong:2000a}, and
rapid onset of optical response, cannot arise from the companion star
alone.

\begin{figure}[t]
\begin{center}
\psfig{width=2.5in,file=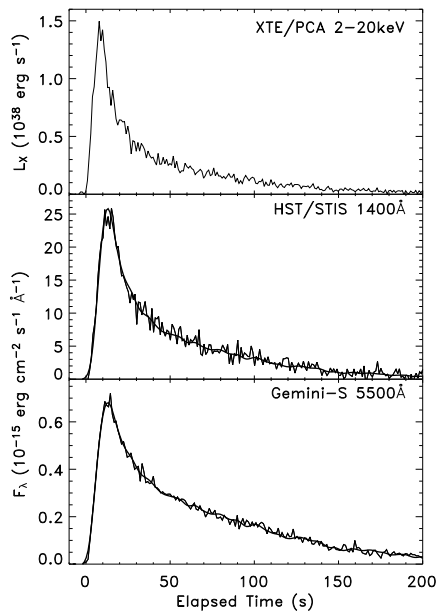}
\end{center}
\caption{Simultaneous burst profiles in EXO~0748--676.  Fits to the
UV/optical data are described in the text.}
\label{BurstFig}
\end{figure}

\subsection{Multiwavelength bursts from EXO~0748--676}

The best dataset for the method yet obtained was obtained for
EXO~0748--676 (\citealt{Hynes:2005a}; see also \citealt{Hynes:2004b}).
This brought together a number of threads from previous successful
studies.  Use of {\it RXTE} provided high quality X-ray coverage,
while a fast CCD camera coupled to the large aperture of the Gemini-S
telescope provided exquisite quality optical lightcurves.  Four bursts
were observed over two successive nights, providing some phase
information, and one of these was also observed at high
time-resolution in the far-UV by {\it HST}/STIS.  The latter was a
unique observation to date, providing far more sensitivity to high
temperature responses than is possible with optical data alone, and
also yielding a time-resolved UV spectrum, facilitating a direct test
of the expectation that the reprocessed light should be close to a
black body.

The burst recorded simultaneously in X-rays, optical, and UV is shown
in Fig.~\ref{BurstFig}.  This was fitted with a model in which the
X-ray burst (expressed as approximately bolometric irradiating
luminosity) was convolved with a Gaussian transfer function to
determine the bolometric reprocessed luminosity.  This was then
converted to $V$ and far-UV fluxes assuming that the reprocessor is a
single-temperature black-body.  This allows superb fit to optical and
UV data, but at the expense of requiring renormalization between the
bandpasses.  The latter may indicate a source of excess optical
reprocessed light from a cooler region (such as the companion star)
which does not contribute to the far-UV.  The best fit parameters
yield a mean lag of about 4\,s and smearing width of 2.5\,s (standard
deviation).  This would be consistent with a combination of emission
from the disk and companion, but is too extended for the disk alone.
The temperature evolution spans the range 18,500--35,200\,K,
comparable to previous measures, but more precisely constrained by the
far-UV coverage.  The far-UV burst spectrum is dominated by the
continuum and is consistent with a slightly reddened 27,000\,K black
body (the flux-weighted average temperature from the light curve fit).
This confirms that the reprocessed burst flux is quite effectively
thermalized.  


\section{Flickering}

While bursts provide an ideal signal for echo-mapping, they also have
limitations.  Not only are bursts not seen in black hole systems, but
there are also neutron star LMXBs which do not burst (e.g.\ Sco X-1).
Clearly we would like complementary echo mapping information about
these systems.  Furthermore, bursts are rather infrequent, with
inter-burst intervals typically 2--10\,hrs.  Accumulating
phase-coverage can therefore be extremely expensive.  An alternative
source of variability is provided by the flickering that seems a
ubiquitous signature of accretion.  While analysis of simultaneous
X-ray and optical/UV flickering can potentially provide phase-resolved
information in any system, this potential has yet to be fully
realized.  It has been found that the optical response is rather weak,
only a few percent at best.  Consequently high signal-to-noise
observations are needed to pick out a measureable correlation.  Even
then, success is typically only achieved when high levels of
variability are present, with other datasets yielding a
non-detection.  For example, in the dataset discussed above on
EXO~0748--676, high quality optical data were obtained, and correlated
bursts were observed, yet no correlation is present between the
inter-burst X-ray and optical lightcurves (other than orbital
variations).

Consequently echo-mapping type analyses based on flickering have only
been published for three objects (excluding XTE~J1118+480 which will
be described separately as a special case).  The bright neutron star
system Sco X-1 was observed by \citet{Ilovaisky:1980a} and
\citet{Petro:1981a}.  Both found correlations, with evidence for lags
and substantial smearing of the response; \citet{Petro:1981a}
described the optical response as a low-pass filtered version of the
X-rays, with variability on timescales $\la20$\,s smoothed out.
\citet{McGowan:2003a} reanalyzed these datasets with the Gaussian
transfer function method.  In some cases no good fit could be
obtained.  The pair of lightcurves where the method did appear to
succeed yielded a lag of $8.0\pm0.8$\,s and Gaussian dispersion of
$8.6\pm 1.3$\,s.  For comparison, lags of up to 4--5\,s are expected
from the disk and 10\,s from the companion star.

The primary focus of \citet{McGowan:2003a} was LMC~X-2, for which new
optical and X-ray data were presented.  A correlation was
detected with the optical data clearly lagging significantly.
Gaussian transfer function fitting yielded a lag of
$18.6^{+7.4}_{-6.6}$\,s with respect to 2--10\,keV X-rays and a
Gaussian dispersion of $10.2^{+5.8}_{-5.7}$\,s.  In the case of this
system, interpretation is less clear as system parameters are not
known.  The orbital period is believed to be 8.2\,hrs (see
\citealt{McGowan:2003a} for discussion).  If the compact object is a
neutron star then we expect disk lags of 2--3\,s and the secondary at
6\,s.  If it is a 10\,M$_{\odot}$ black hole then disk and companion
lags of 4--5\,s and 11\,s are expected respectively.  Thus for both
Sco~X-1 and LMC~X-2 there is marginal evidence for a response
extending to later than expected from light travel times alone. 

Finally the black hole system GRO~J1655--40 was observed
simultaneously by {\it HST} and {\it RXTE} on several occasions.  On
one of these clear correlations were seen between rapid X-ray and
optical variability \citep{Hynes:1998a}.  These were repeatable
across multiple independent pairs of lightcurves and implied lags of
10--20\,s with smearing times of order 10\,s.  This system has well
determined orbital parameters allowing direct comparison of the
measured lags with expectations.  At the orbital phase observed the
companion star should have produced lags of over 40\,s so is ruled out
by causality.  The lags measured are consistent with being
dominated by the disk.  An analysis using modeled binary transfer
functions by \citet{OBrien:2002a} came to the same conclusion.

\begin{figure}[t]
\begin{center}
\psfig{angle=90,width=5in,file=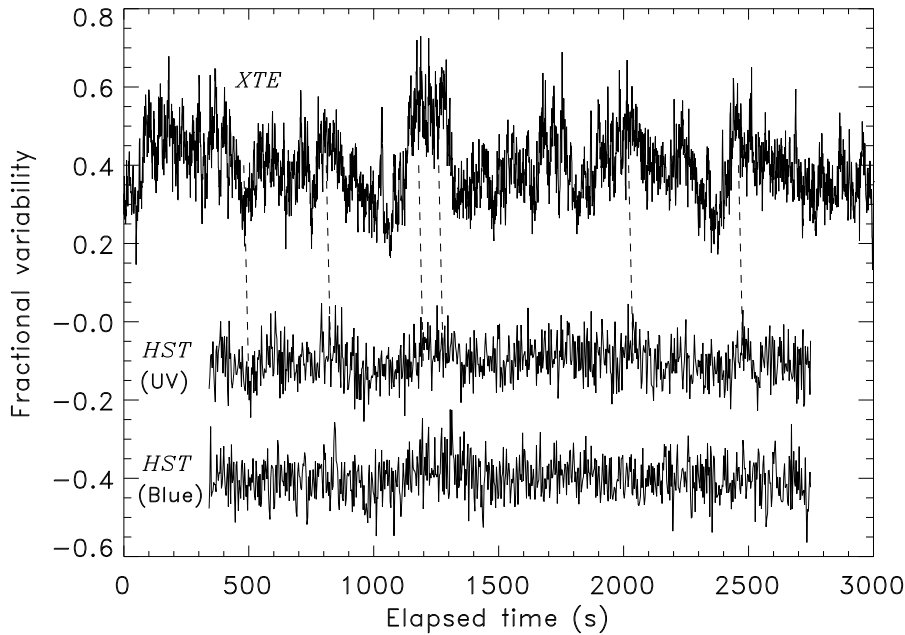}
\end{center}
\caption{Simultaneous lightcurves from the black hole system
  GRO~J1655--40.  Correlated variations are clearly present and a lag
  can just about be seen directly in the lightcurves.  From
  \citet{Hynes:1998a}.}
\label{HSTFig}
\end{figure}


\section{Echoes in Quiescence}

Thus far we have only discussed echo-mapping experiments in the
context of X-ray bright states.  Although optical flickering is
present in quiescent X-ray binaries (e.g.\ \citealt{Zurita:2003a};
\citealt{Hynes:2003a}), it has generally been assumed that the X-rays
are too faint to cause measureable reprocessing and that optical
flickering arises from similar mechanisms as in cataclysmic variables,
in the outer disk and/or stream-impact point.  The highest quality
optical observations reveal lightcurves clearly distinct from those of
cataclysmic variables, however, calling this assumption into question.
The strongest evidence against it came from spectroscopic observations
of optical flares in the black hole system V404~Cyg
\citep{Hynes:2002a}.  Emission line flares exhibited the
characteristic double-peaked profile of emission distributed across an
accretion disk, leading to the speculation that they might be global
events driven by irradiation rather than local magnetic reconnection
events.  This was strikingly confirmed by simultaneous X-ray and
optical observations of the system demonstrating a clear correlation
between the X-rays and both the optical line emission (which was again
double-peaked) and the continuum \citep{Hynes:2004a}.  The flare
timescales, and lags between X-rays and optical, are much shorter than
the dynamical timescales in the disk regions implied by the line
profile enhancements, indicating that only light travel timescales can
couple these regions fast enough.

\begin{figure}[t]
\begin{center}
\psfig{angle=90,width=5in,file=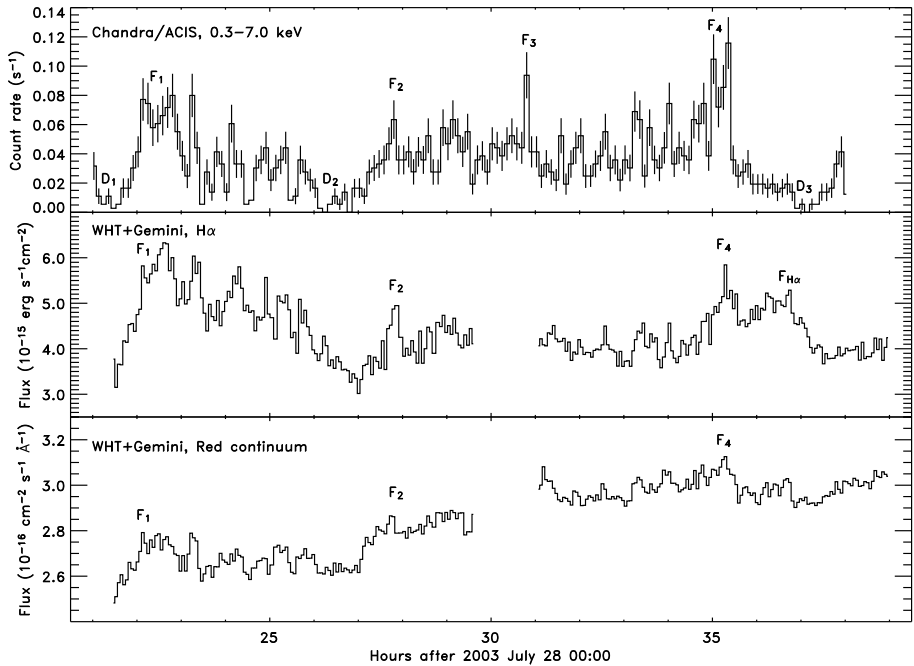}
\end{center}
\caption{Simultaneous {\em quiescent} lightcurves of the black hole
  system V404~Cyg at X-ray and optical wavelengths.  Note the
  correlation between X-ray and optical variability.  The lines must
  be formed through reprocessing, although the origin of the continuum
  remains unclear.  From \citet{Hynes:2004a}.}
\label{ChandraFig}
\end{figure}

This discovery raises the possibility of applying echo-mapping (and
even echo-tomography) to determine the geometry of the accretion flow
in quiescent systems.  Furthermore, the strong line response allows us
to obtain kinematic information on the reprocessing site as well as
light travel time diagnostics.  The existing observations rule out
large X-ray to line lags, but shorter lags consistent with the
expected light travel time across the disk ($\sim40$\,s) are allowed.
A preliminary analysis suggests the lag may be even larger than this,
raising again the specter of diffusion times.  As discussed earlier,
these might be expected to be large for the cool material expected in
quiescent disks (\citealt{Cominsky:1987a}; \citealt{McGowan:2003a}).
If this effect is important then it may compromise the utility of
echo-mapping in these systems, although its presence would instead provide
diagnostic information on the physical conditions in the reprocessing
regions. 


\section{When is an Echo Not an Echo?}

Fundamental to echo-mapping is the assumption that correlated X-ray
and optical/UV variability indicate reprocessing of the X-rays by
relatively cool material.  We should not take this for granted,
however, and there are some observations which seriously challenge
this assumption.

The first indication of difficulties came from fast optical
observations of the black hole binary GX\,339--4
(\citealt{Motch:1982a}; \citealt{Motch:1983a}).  Dramatic optical
variability was seen extending to extremely short timescales
(10--20\,ms), much shorter than the light travel timescales expected,
or the smearing typically observed in other systems described above.
\citet{Fabian:1982a} argued that the flares most likely originated
cyclotron radiation, with a brightness temperature
$\ga9\times10^8$\,K.  Correlations were seen in a short (96\,s)
simultaneous observation, but of a puzzling nature.  The X-ray and
optical were anti-correlated with optical dips apparently preceding the
X-rays by a $2.8\pm1.6$\,s.  The connection
between X-ray and optical behavior was further reinforced by the
presence of quasi-periodic oscillations at the same frequency in both
energy bands.  The brevity of the simultaneous observation, and the
ambiguity in the lags introduced by quasi-periodic variability left
this result tantatlizing however.

New light was shed on this behavior by the 2000 outburst of the black
hole system XTE~J1118+480.  A much larger time-resolved database was
accumulated on this object including both multi-epoch simultaneous
X-ray/UV observations (\citealt{Haswell:2000a}; \citealt{Hynes:2003b})
and independent X-ray/optical data (\citealt{Kanbach:2001a};
\citealt{Spruit:2002a}; \citealt{Malzac:2003a}).  Large amplitude
X-ray variability was present, and accompanied by correlated UV and
optical variations.  In this case a positive correlation with the
optical/UV lagging the X-rays was clearly present leading to hopes
that this would be an ideal echo-mapping dataset.  There were serious
problems with this interpretation, however.  These were most
pronounced in the optical data \citep{Kanbach:2001a} and included an
optical auto-correlation function narrower than that seen in X-rays,
and a cross-correlation function containing a marked ``precognition
dip'' before the main peak.  The latter could be interpreted in terms
of optical dips leading X-ray flares by a few seconds, as suggested in
GX\,339--4, suggesting a common origin.  Neither of these effects are
expected in a reprocessing model.  Light travel times should only act
to smooth out optical responses, and hence broaden the optical
auto-correlation function, and continuum responses (as considered
here) should generally be positive.  As in GX\,339--4 the variability
extended to very short timescales ($\la100$\,ms) and hence
\citet{Kanbach:2001a} estimated a minimum brightness temperature of
$2\times10^6$\,K.  They also suggested that the strange variability
properties were the result of optical cyclosynchrotron emission. These
properties become weaker at shorter wavelengths \citep{Hynes:2003a},
as does the variability, as might be expected if the behavior
originates from a very red source of emission like synchrotron.

Dominant synchrotron emission in this system was not uniquely
suggested by the variability properties.  The very flat UV to near-IR
spectrum had previously been attributed to synchrotron emission
\citep{Hynes:2000a} and the broad-band spectral energy distribution
has been successfully accounted for using a simple jet model
\citep{Markoff:2001a}.  In the latter model, only the IR originates
from flat-spectrum (self-absorbed) synchrotron, whereas the extension
of the flat spectrum into the UV is explained as a coincidental
combination of optically thin synchrotron and disk emission.  The
variability amplitudes define a rather different energy distribution
to that seen in persistent light.  The variability decreases from
nearly 50\,\%\ rms in the $K$ band to just a few percent in the far-UV
\citep{Hynes:2003b}.  When these fractional amplitudes are converted
into absolute flux units, the rms variability energy distribution
defines a power-law with $F_{\nu} \propto \nu^{-0.59}$, totally
consistent with optically thin synchrotron.  Furthermore, this
power-law not only applies to the IR--UV variability, but extends to
X-rays, and accounts for both the amplitude and energy dependence of
X-ray variability.  This implication of synchrotron X-ray emission
remains controversial, but was also predicted by the jet model of
\citet{Markoff:2001a}.  Recently \citet{Malzac:2004a} have shown that it is
also possible to reproduce the auto-correlation and cross-correlation
function properies using a model in which a common magnetic reservoir
drives both X-ray variations in an accretion flow (in this case
assumed to be inverse Compton emission rather than synchrotron) and
optical variability in a jet.

These arguments together provide strong evidence that a jet, or at
least some kind of outflow, is responsible for much of the emission in
XTE~J1118+480 and for the correlated variability.  By extension, the
same interpretation may apply to other objects showing similar
properties.  GX\,339--4 is of course a prime candidate, but it is also
argued that jets may be dominant in quiescent black hole systems
\citep{Fender:2003a}.  It may thus be that the {\em continuum}
component correlated with X-rays in V404~Cyg in quiescence actually
arises from jet variability, rather than reprocessing as required to
explain the line emission.


\section{Conclusion}

Correlated X-ray and optical/UV variability has now been widely, if
not commonly, detected in X-ray binaries.  The detections span the
lowest luminosities to the highests and suggest a more complex origin
than the simple reprocessing originally envisioned.  

Reprocessing does appear to be responsible for reprocessed type I
X-ray bursts, and for correlated flickering in luminous X-ray
binaries.  Typically these studies have shown that the disk appears to
dominate the response, although in some cases the lags appear to lag
for light travel times alone.  Where a reprocessed spectrum has been
obtained it is consistent with predominantly optically thick thermal
reprocessing.  The full potential of this technique has yet to be
exploited, either through phase-resolved echo-tomography or through
kinematically resolved emission line echo-mapping.

On the other hand, correlated variations in X-ray binaries in the
low/hard state do not have the characteristics of reprocessed
variability, and instead appear to originate in optical synchrotron
emission, likely from a jet.  Understanding the variability properties
in the context of a jet model remains a topic of ongoing research, but
this may prove a very valuable diagnostic of the disk-jet connection.

At the lowest luminosities, there is evidence that emission line
variability in V404~Cyg in quiescence is dominated by reprocessing
within the disk.  The origin of continuum variability is less clear,
and this might either be reprocessing, or synchrotron.  For these
systems the low X-ray brightnesses are a challenge that Chandra and
XMM are barely adequate for, and full exploitation of echo-mapping in
quiescence will likely require a larger throughput mission such as
Constellation-X. 


\acknowledgements{I would like to acknowledge the contributions of my
  collaborators in this field; Carole Haswell, Keith Horne, and Kieran
  O'Brien deserve special mention alongside many others too numerous
  to name.  I am grateful for support from NASA through Hubble
  Fellowship grant \#HF-01150.01-A awarded by STScI, which is operated
  by AURA, for NASA, under contract NAS 5-26555. This work has made
  extensive use of the NASA ADS Abstract Service.}


\end{document}